\documentclass[useAMS,usenatbib]{mn2e}
\usepackage{epsfig}
\usepackage{epstopdf}
\usepackage{amsmath}
\usepackage{amssymb}
\usepackage{natbib}
\newcommand{\apj}{ApJ}
\newcommand{\apjl}{ApJ}
\newcommand{\mnras}{MNRAS}

\newcommand{\nat}{Nat}
\newcommand{\araa}{ARAA}
\newcommand{\pasp}{PASP}

\newcommand{\physrep}{Physics Reports}

\newcommand{\microns}{\micron}

\newcommand{\cm}{\ensuremath{{\rm cm}}}
\newcommand{\km}{\ensuremath{{\rm km}}}
\newcommand{\pc}{\ensuremath{{\rm pc}}}
\newcommand{\kpc}{\ensuremath{{\rm kpc}}}

\newcommand{\yr}{\ensuremath{{\rm yr}}}

\newcommand{\erg}{\ensuremath{{\rm erg}}}
\newcommand{\K}{\ensuremath{{\rm K}}}
\newcommand{\s}{\ensuremath{{\rm s}}}

\newcommand{\Td}{\ensuremath{T_{\rm dust}}}
\newcommand{\Sg}{\ensuremath{\Sigma_{\rm g}}}

\newcommand{\lsun}{\ensuremath{{\rm L_{\odot}}}}
\newcommand{\msun}{\ensuremath{{\rm M_{\odot}}}}

\newcommand{\mgas}{\ensuremath{{M_{\rm gas}}}}
\newcommand{\ngas}{\ensuremath{{n_{\rm gas}}}}

\newcommand{\fmol}{\ensuremath{f_{\rm H_{2}}}}
\newcommand{\fco}{\ensuremath{f_{\rm CO}}}
\newcommand{\fcii}{\ensuremath{f_{\rm CII}}}

\newcommand{\lcii}{\ensuremath{L_{\rm [CII]}}}
\newcommand{\lfir}{\ensuremath{L_{\rm FIR}}}

\newcommand{\ssfr}{\ensuremath{{\dot{\Sigma}_{\rm \star}}}}

\usepackage{wasysym}
\usepackage[usenames]{color}
\definecolor{dkgreen}{RGB}{0,200,0}

\begin{document}

\title[The {[CII]} Deficit and High-Temperature Saturation]{The [CII] Deficit in LIRGs and ULIRGs is Due to High-Temperature Saturation}

\author[Mu{\~n}oz \& Oh]{
Joseph~A.~Mu{\~n}oz\thanks{jamunoz@physics.ucsb.edu}
and
S.~Peng~Oh\thanks{peng@physics.ucsb.edu}\\
University of California Santa Barbara, Department of Physics, Santa Barbara, CA 93106, USA\\
}

\maketitle

\begin{abstract}
Current predictions for the line ratios from photo-dissociative regions (PDRs) in galaxies adopt theoretical models that consider only individual parcels of PDR gas each characterized by the local density and far-UV radiation field.  However, these quantities are not measured directly from unresolved galaxies, making the connection between theory and observation ambiguous.  We develop a model that uses galaxy-averaged, observable inputs to explain and predict measurements of the [CII] fine structure line in luminous and ultra-luminous infrared galaxies.  We find that the [CII] deficit observed in the highest IR surface-brightness systems is a natural consequence of saturating the upper fine-structure transition state at gas temperatures above $91\,\K$.  To reproduce the measured amplitude of the [CII]/FIR ratio in deficit galaxies, we require that [CII] trace approximately 10--17$\%$ of all gas in these systems, roughly independent of IR surface brightness and consistent with observed [CII] to CO(1--0) line ratios.  Calculating the value of this fraction is a challenge for theoretical models.  The difficulty may reside in properly treating the topology of molecular and dissociated gas, different descriptions for which may be observationally distinguished by the [OI]$63\,\micron$ line in yet-to-be-probed regions of parameter space, allowing PDR emission lines from to probe not only the effects of star formation but also the state and configuration of interstellar gas.
\end{abstract}


\section{Introduction}

The strength and ubiquity of the [CII] fine structure emission line makes it an important observational diagnostic of star formation in galaxies.  However, while emission from fine structure lines in photo-dissociative regions (PDRs) of galaxies have been extensively modeled theoretically \citep[e.g.,][]{TH85, Wolfire90, Hollenbach91, Kaufman99}, these calculations consider individual parcels of PDR gas parameterized by the local gas density and radiation field, inputs that are not readily measurable in distant galaxies.  Thus, observations are rarely used to test the theoretical predictions.  Rather, the gas density and radiation field are deduced in some average sense by comparing spatially-unresolved emission lines to the models and requiring agreement.

Moreover, for luminous and ultra-luminous infrared galaxies (LIRGs and ULIRGs), the [CII] line ceases to be a reliable empirical calorimeter of star formation.  Instead a [CII] `deficit' is observed \citep[e.g.,][]{Malhotra97, Luhman98, Malhotra01, Luhman03}.  While this result is quantified by a number of relationships among semi-correlated observables, perhaps the most reasonable description is as a decreasing ratio of [CII]/FIR with increasing IR surface brightness \citep{Diaz-Santos13, Diaz-Santos14}, since IR surface brightness, which comprises both the energy injection by star formation and the compactness of a region, traces the heating rate in the interstellar medium (ISM).  The deficit also becomes more pronounced with an increasing ratio of $60\,\micron$ to $100\,\micron$ continuum flux, suggesting some relationship to the dust temperature \citep[e.g.,][see also \citealt{Gullberg15}]{Malhotra97, Malhotra01, Diaz-Santos13}.\footnote{\citet{Diaz-Santos13} replaced the commonly used FIR color $S_{\rm 60\,\microns}/S_{\rm 100\,\microns}$ with $S_{\rm 63\,\microns}/S_{\rm 158\,\microns}$ both to probe a wider range of dust temperatures over a longer wavelength separation and to make use of full resolution PACS images, enabling them to distinguish between galaxies otherwise blended in the {\it{IRAS}} data.}  Moreover, a similar behavior is observed for other PDR emission lines as well \citep{Gracia-Carpio11}.  

Several different explanations for this measured decline abound, including reduced photoelectric heating owing to positively-charged dust grains and absorption of ionizing and UV photons by dust in HII regions \citep[see][for a recent review]{Casey14}, but these analyses seem to ignore the simplest and most obvious solution: the saturation of the upper energy level at gas temperatures much larger than that corresponding to the energy difference of the transition.  \citet{Luhman03} briefly hint at this solution, attributing it to \citet[][though these authors actually advocate inefficient photoelectric heating]{GC01}, but dismiss it because of a presumed requirement for densities of $\sim\!\!10^{5}\,\cm^{-3}$.  [CII] saturation at high temperatures has also been mentioned as a possibility by other authors \citep[e.g.,][]{Stacey10, Diaz-Santos13, Gullberg15}, who do not make detailed theoretical predictions for the [CII] deficit.

\citet{Luhman03} further point out the importance of the relative filling factors of FIR and [CII] in using the ratio of the two as a probe of the ISM; the FIR emission is optically thick and comes from all gas containing dust, while the [CII] emission is optically thin but originates only from dissociated gas.  However, the fraction of dissociated gas is challenging to compute theoretically.  For example, \citet{MF14} underestimated the [CII] emission by about an order-of-magnitude using the \citet{Krumholz11} molecular fraction and the \citet{Wolfire10} dark molecular gas fraction models to compute the gas mass fraction traced by [CII].

To explore the high-temperature saturation mechanism for the [CII] deficit and understand how it probes galaxy and PDR properties, we combine local calculations of the [CII] line cooling rate in PDRs with simple average galaxy models.  In \S\ref{sec:deficit}, we decompose the [CII]/FIR ratio of a galaxy into small- and large-scale factors.  We then examine each component: the specific [CII] luminosity, the fraction of gas traced by [CII], and the specific FIR luminosity in \S\ref{sec:cii}, \S\ref{sec:fcii}, and \S\ref{sec:fir}, respectively.  In \S\ref{sec:results}, we demonstrate the excellent agreement between recent observations and our predictions for the relationships between [CII]/FIR and either IR surface brightness or dust temperature.  We review two other popular mechanisms to explain the [CII] deficit in \S\ref{sec:alt} and compare them to saturated [CII] emission in the high-temperature limit.  Finally, we conclude in \S\ref{sec:conc}.

\section{Decomposing the [CII] Deficit}\label{sec:deficit}

There is a prevalent ambiguity in the literature about whether the [CII] deficit probes local behavior in PDRs or is a global property of the galaxy as a whole.  To clarify this question, we express the ratio of [CII] to FIR luminosity as
\begin{equation}\label{eq:ratio}
\frac{\lcii}{\lfir}=\frac{\lcii}{\mgas}\,\frac{\mgas}{\lfir}=\frac{\lcii}{M_{\rm CII}}\,\fcii\,\left(\frac{\Sigma_{\rm FIR}}{\Sg}\right)^{-1},
\end{equation}
where $\lcii$ is the [CII] luminosity, $\lfir$ is the FIR luminosity, $\fcii=M_{\rm CII}/\mgas$ is the ratio of the gas mass containing dissociated carbon to the total gas mass in the galaxy, and $\Sigma_{\rm FIR}$ and $\Sg$ are the FIR surface brightness and surface density of the galaxy, respectively.   This decomposition demonstrates that the [CII] ratio results from a mix of small- and large-scale effects.  In equation~\ref{eq:ratio}, the specific [CII] luminosity, $\lcii/M_{\rm CII}$, is a local quantity in regions containing dissociated carbon, while the specific FIR luminosity is averaged over larger scales, either over many star-forming regions or over a whole galaxy.

\section{The Specific [CII] Luminosity}\label{sec:cii}

To compute the specific [CII] luminosity in equation~\ref{eq:ratio}, consider a gas cloud of temperature $T_{\rm gas}$ and density $\ngas$ higher than $\sim\!\!10^{3}\,\cm^{-3}$, the critical density for thermalization of the $158\,\micron$ fine structure transition, in which a solar abundance of carbon (${\rm C/H}=1.1\times10^{-4}$) is entirely in the form of dissociated CII.  If the CII itself is optically thin to its own emission, then the resulting cooling rate per hydrogen atom through the [CII] line is
\begin{equation}\label{eq:Lcii}
\Lambda_{{\rm [CII], thermal}}=A_{\rm [CII]}\,k_{\rm B}\,T_{\rm [CII]}\,\frac{2}{{\rm e}^{T_{\rm [CII]}/T_{\rm gas}}+2}\,\frac{\rm C}{\rm H},
\end{equation}
where $h$ and $k_{\rm B}$ are the Planck and Boltzmann constants, respectively, $A_{\rm [CII]}=2.3\times10^{-6}\,\s^{-1}$ is Einstein emission coefficient, and $T_{\rm [CII]}=h\,\nu_{158\,\microns}/k_{\rm B}\approx 91.25\,\K$ is transition temperature.  As the gas temperature approaches infinity, the Boltzmann factor in the denominator of equation~\ref{eq:Lcii} goes to unity, and thermal [CII] emission per gas mass saturates at
\begin{equation}\label{eq:Lcii1}
\frac{\lcii}{M_{\rm CII}}=0.66\,\frac{\lsun}{\msun}.
\end{equation}
The same mechanism results in saturation for other PDR emission lines as well, which may be influential in the corresponding observed deficits for those lines \citep{Gracia-Carpio11}.  Additionally, note that a similar high-temperature saturation also occurs if the gas density is constant but below the critical value, since the emission again depends on temperature primarily (though not exclusively, in this case) through the Boltzmann factor.

Because of the saturation of [CII], the [OI]$63\,\micron$ line begins to dominate the cooling until it too saturates at temperatures beyond $T_{\rm [OI]63}\approx227.7\,\K$ and gas densities above a critical value of about $10^{5}\,\cm^{-3}$.  Below such high densities, the subthermal cooling rate per hydrogen atom is
\begin{equation}\label{eq:Loi}
\Lambda_{{\rm [OI]63, subthermal}}=\frac{3}{5}\,k_{\rm [OI]63}\,\ngas\,k_{\rm B}\,T_{\rm [OI]63}\,\frac{\rm O}{\rm H},
\end{equation}
where the collisional excitation rate is $k_{\rm [OI]63\text{--}H}\approx 3.6\times10^{-10}\,\s^{-1}\,\cm^{3}\,{\rm e}^{-T_{\rm [OI]63}/T}$ for collisions with atomic hydrogen\footnote{http://home.strw.leidenuniv.nl/{\raise.17ex\hbox{$\scriptstyle\sim$}}moldata/O.html} and ${\rm O}/{\rm H}\approx 5.0\times10^{-5}$.  Because the [CII] and [OI]$63\,\micron$ lines dominate the cooling rate in PDRs \citep[e.g.,][]{TH85}, equating the combined cooling rate from equations~\ref{eq:Lcii} and~\ref{eq:Loi} with the photo-electric heating rate in equation~\ref{eq:Hpe} given by \citet{Wolfire03} yields the equilibrium temperature as a function of $\ngas$ and the $G'$, the FUV radiation field scaled to the local solar value.  Conservatively adopting a low ratio of $G'/\ngas$ for LIRGs, with values of $\ngas=10^{4}\,\cm^{-3}$ and $G'=100$ \citep[see, e.g.,][]{Luhman03}, gives an equilibrium gas temperature of $T_{\rm gas}\approx 106\,\K$, which is in the saturated [CII] regime.  The equilibrium temperature reaches even higher values for the higher ratios of $G'/\ngas$ more appropriate for LIRGs and ULIRGs.  This justifies our assumption of high temperature in equation~\ref{eq:Lcii1}.

\section{The CII Fraction}\label{sec:fcii}

In the high-temperature and density limit of equation~\ref{eq:Lcii1}, [CII] emission is a direct tracer of the gas mass in which carbon is dissociated.  To calculate the fraction of the total mass represented by this [CII]-traced component, note that the remaining mass is probed by the CO emission.  A value of $\alpha_{\rm CO}=1\,\msun\,\pc^{-2}\,(\K\text{--}\km\,\s^{-1})^{-1}$ for the ratio of the molecular mass to CO(1--0) luminosity---roughly appropriate for dusty, star-forming galaxies \citep{DS98}---indicates an emission rate of $2\times10^{4}\,\lsun$ in the CO(1--0) line per solar mass of molecular gas.  Combining this with equation~\ref{eq:Lcii1} yields the fraction of total gas traced by CII approximately as
\begin{equation}\label{eq:fcii_obs}
\fcii \approx \frac{1}{1+3\times10^{4}\,\alpha_{1}\,L_{\rm CO(1-0)}/\lcii},
\end{equation}
where $\alpha_{1}=\alpha_{\rm CO}\,\msun^{-1}\,\pc^{2}\,(\K\text{--}\km\,\s^{-1})^{1}$.

Empirically, the ratio of [CII] to CO(1--0) luminosity is a roughly constant value of 4400 among both Milky Way and extragalactic sources spanning a range of temperatures, densities, luminosities, masses, and redshifts \citep[e.g.,][]{Crawford85, Wolfire89, Stacey91, Stacey10}.  Using equation~\ref{eq:fcii_obs}, this ratio implies $\fcii\approx 0.13$.  Within a recent high-redshift sample from \citet{Gullberg15} the ratio, similarly, is $5200\pm1800$ corresponding to $\fcii \approx 0.15\pm0.04$.

There are a couple of caveats to the above calculation.  First, equation~\ref{eq:fcii_obs} assumes that the density and temperature of the [CII] emitting gas are well above $10^{3}\,\cm^{-3}$ and $91\,\K$, respectively.  If not in this limit, $\fcii$ would decrease, though \S\ref{sec:cii} demonstrates that galaxies exhibiting a [CII] deficit are likely in the high-temperature regime.  Note that the high-temperature and density limit for [CII] emission is also assumed by observational estimates of the dissociated to molecular mass ratio in the literature \citep[e.g.,][]{Stacey10}.

Additionally, equation~\ref{eq:fcii_obs} double-counts the contribution from `dark' molecular gas in which the hydrogen is sufficiently self-shielded to be in molecular form but is still susceptible to CO-dissociating radiation and, thus, emits in [CII].  This double counting occurs because both $\fcii$ and $\alpha_{\rm CO}$ include dark H$_{2}$.  To correct for this, we should multiply $\alpha_{\rm CO}$ by an additional factor of $(1-f_{\rm DG}/\fmol)$, where $\fmol$ and $f_{\rm DG}$ are the molecular and dark molecular gas fractions as defined in \citet{Wolfire10}.  However, the low value of $\alpha_{\rm CO}$ adopted here as appropriate for LIRGs and ULIRGs implies that the dark fraction is small, which is consistent with theoretical calculations at high surface densities \citep{Wolfire10, Narayanan12, MF14}.  Moreover, the fractional change in $\fcii$ accounting for the over-counted dark fraction is approximately $f_{\rm DG}/\fmol$.  For example, if $f_{\rm DG}/\fmol$ is as high as 0.3, our estimate of $\fcii$ increases only about $30\%$, from 0.13 to 0.17.

Finally, equation~\ref{eq:fcii_obs} ignores gas in which the carbon may be in the form of CI rather than in CO or CII.  A significant amount of such gas would lower our estimate for $\fcii$.  However, CI is the dominant form of carbon in parts of a PDR only if the gas density is quite low \citep[$\sim\!\!10^{2}\,\cm^{-2}$,][]{Hollenbach91}, below the critical value for [CII].

A constant value of 10--20$\%$ for $\fcii$ is difficult to derive theoretically.  For model galaxies comparable to the sub-millimeter galaxy HFLS3 \citep{Riechers13} and the quasar hosts J2310$+$1855 \citep{Wang13} and J1148$+$5251 \citep{Riechers09}, combining the \citet{Krumholz11} molecular fraction and the \citet{Wolfire10} `dark' molecular gas fraction models yields a value of $\fcii\sim0.01$, under-estimating the observed [CII] emission by roughly an order-of-magnitude \citep{MF14}.  Moreover, $\fcii$ is quite sensitive to galaxy surface density in these models \citep{MF13, MF14}, which is inconsistent with the relatively constant observed values.  Note, however, that an order-of-magnitude increase in $\fcii$ from 0.01 to 0.1 decreases $\fco$ by only about $10\%$---from 0.99 to 0.9---suggesting that derivations of the CO luminosity based on this modeling \citep[e.g.,][]{Krumholz11, Narayanan12} are likely not significantly affected.

One possible issue with the above approach to calculating $\fcii$ is that the \citet{Wolfire10} model assumes clouds of discrete, shielded molecular gas bathed in an interstellar background, which may not be an appropriate description of the ISM in LIRGs and ULIRGs.  Rather, high surface density systems are more likely represented by a `swiss cheese' picture in which relatively isolated pockets of PDR gas surround the HII regions of star clusters in a mostly molecular medium.  In this case, the PDR gas is subject to the local cluster radiation field, which would be larger but more invariant from galaxy to galaxy than the large-scale background field.  Such a scenario may, thus, be observationally distinguished by temperature-dependent, [OI]$63\,\micron$ line emission and represent a way forward for analytic calculations to complement future high-resolution numerical simulations of star formation and HII regions in galaxies.

\section{The Specific FIR Luminosity}\label{sec:fir}

Unlike the [CII] emission, the IR luminosity traces the total energy injected by star formation.  To describe the reprocessing into the IR, assume that the dust is very optically thick with a fixed fraction $f_{\rm IR}$ of the total bolometric energy from star formation being emitted in the IR.  The IR surface brightness is then given by 
\begin{equation}\label{eq:SIR}
\begin{split}
\Sigma_{\rm IR} &\approx \frac{1}{2}\,f_{\rm IR}\,\epsilon\,\ssfr\,c^2\\
&\approx 0.6\times10^{10}\,\lsun\,(\msun/\yr)^{-1}\,\left(\frac{f_{\rm IR}\,\epsilon}{0.8\times10^{-3}}\right)\,\ssfr,
\end{split}
\end{equation}
where $\epsilon \approx 10^{-3}$ and $f_{\rm IR}\approx 0.8$ are set based on stellar population synthesis modeling and recover the usual conversion between IR luminosity and star formation rate adopted in the literature \citep[e.g.,][]{Kennicutt98, Murphy11}.
Further assuming that half of the IR emission falls into the FIR, the resulting ratio of FIR surface brightness to gas surface density is
\begin{equation}\label{eq:FIRthick}
\frac{\Sigma_{\rm FIR}}{\Sg} \approx 3\times10^{3}\,\frac{\lsun}{\msun}\,\left(\frac{\ssfr}{\msun\,\yr^{-1}\,\kpc^{-2}}\right)\,\left(\frac{\Sg}{\msun\,\pc^{-2}}\right)^{-1},
\end{equation}
where, again, $\ssfr$ and $\lfir/\mgas$ are values averaged over many star-forming regions or over a whole galaxy.  

Over these large scales, the surface star formation rate and surface gas density are empirically related by the observed star formation law, which, when corrected for a varying CO X-factor, is roughly given by \citep[see][]{OS11}:
\begin{equation}\label{eq:KS}
\ssfr \approx 9.2\,\left(\frac{\Sg}{10^3\,\msun\,\pc^{-2}}\right)^2\,\msun\,\yr^{-1}\,\kpc^{-2},
\end{equation}
which holds for surface densities above about $100\,\msun/\pc^2$.  Note that this result can also be derived theoretically by considering the hydrodynamical balance between gravitational collapse and turbulent pressure support in marginally Toomre-stable regions \citep[e.g.,][]{Thompson05, OS11}.  Combining equations~\ref{eq:SIR}, \ref{eq:FIRthick} and~\ref{eq:KS} yields
\begin{equation}\label{eq:SFIR}
\begin{split}
\frac{\lfir}{\mgas} =\frac{\Sigma_{\rm FIR}}{\Sg} &\approx 30\,\frac{\lsun}{\msun}\,\left(\frac{\ssfr}{10\,\msun\,\yr^{-1}\,\kpc^{-2}}\right)^{1/2}\\
&\approx 40 \,\frac{\lsun}{\msun}\,\left(\frac{\Sigma_{\rm IR}}{10^{11}\,\lsun\,\kpc^{-2}}\right)^{1/2}
\end{split}
\end{equation}
for the last term on the right-hand-side of equation~\ref{eq:ratio}.  

The surface star formation rate additionally quantifies the effective temperature of the dust, $\Td$, in the galaxy via the bolometric equivalent to equation~\ref{eq:SIR}:
\begin{equation}\label{eq:Tdust}
\Td^4 = \frac{\epsilon\,\ssfr\,c^2}{2\,\sigma_{\rm SB}} \approx (15\,{\rm K})^4\,\frac{\ssfr}{\rm \msun/yr/kpc^2},
\end{equation}
with $\sigma_{\rm SB}$ the Stefan-Boltzmann constant and ignoring heating from the cosmic microwave background as in equation~\ref{eq:SIR}.  Considering equations~\ref{eq:SIR}, \ref{eq:SFIR}, and~\ref{eq:Tdust}, it is clear that the surface star formation rate, $\ssfr$---which comprises the energy injection rate from stars and the compactness of the region in which they form---is the key quantity determining the behavior of all of the evolving quantities relevant to the [CII] deficit.

\section{Results}\label{sec:results}

\begin{figure}
\begin{center}
\includegraphics[width=\columnwidth,trim=5 30 5 0,clip]{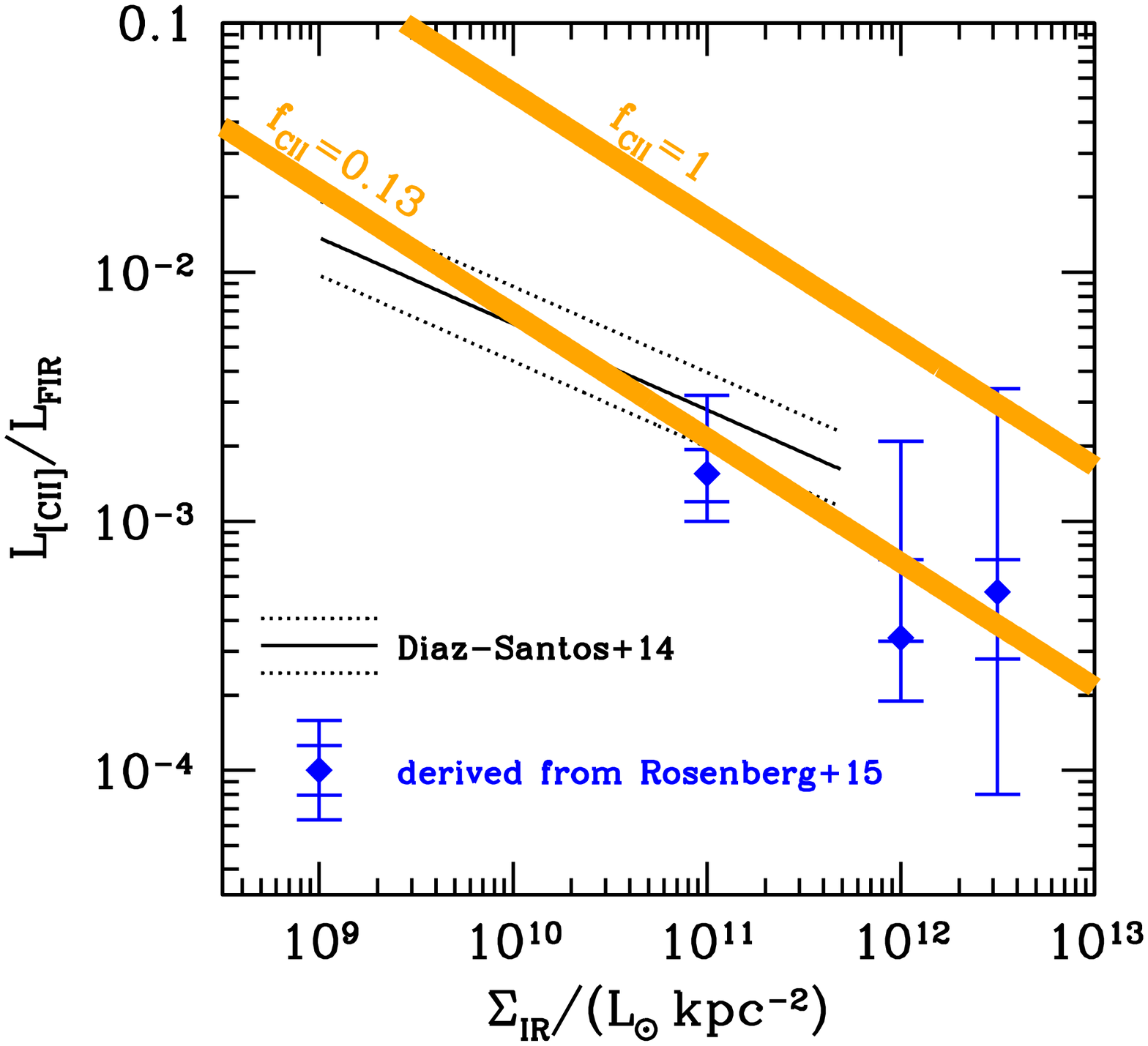}
\caption{\label{fig:ratio} 
The ratio of [CII] to FIR as a function of IR surface brightness.  Thick (orange) curves show results from equation~\ref{eq:ciidef} for $\fcii=0.13$ and 1.  The thin, solid and dotted (black) lines give the fit to the mean and spread of observations from \citet{Diaz-Santos14}.  Points (blue) show the median values for the three classes outlined in \citet{Rosenberg15} with values of $\Sigma_{\rm IR}$ derived from comparing average CO SLEDs to the theoretical SLED model of \citet{NK14}.  Error bars on these points denote the central $50\%$ and maximum and minimum values of each group.}
\end{center}
\end{figure}

\begin{figure}
\begin{center}
\includegraphics[width=\columnwidth,trim=5 30 5 0,clip]{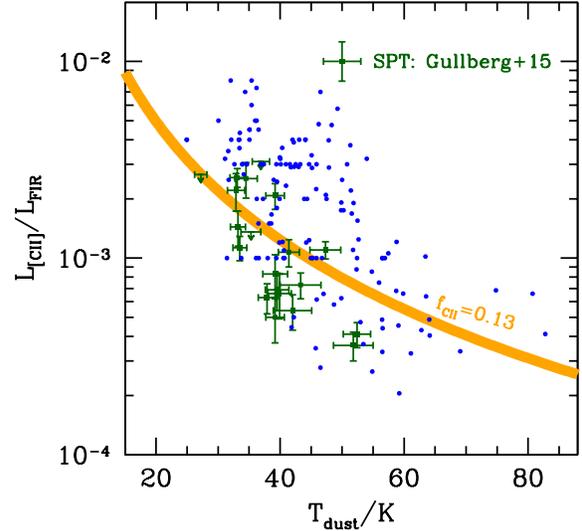}
\caption{\label{fig:ratio1} 
The ratio of [CII] to FIR as a function of dust effective temperature.  The thick (orange) curve presents results from equation~\ref{eq:ciidef1} for $\fcii=0.13$, while square points (green) and error bars show South Pole Telescope (SPT) observations from \citet{Gullberg15}.  For comparison, circular points (blue) show unpublished, low-redshift data from a Great Observatories All-Sky Survey (GOALS) sample (Graci\'{a}-Carpio et al., in preparation).}
\end{center}
\end{figure}

In \S\ref{sec:deficit}, we decomposed the [CII] deficit into the specific [CII] luminosity, the fraction of gas traced by [CII], and the specific FIR luminosity and subsequently derived each of these quantities in \S\ref{sec:cii}, \S\ref{sec:fcii}, and \S\ref{sec:fir}, respectively.  Now, substituting equations~\ref{eq:Lcii1} and~\ref{eq:SFIR} back into equation~\ref{eq:ratio} and scaling $\fcii$ to 0.13, yields
\begin{equation}\label{eq:ciidef}
\frac{\lcii}{\lfir} \approx 2.2\times10^{-3}\,\frac{\fcii}{0.13}\,\left(\frac{\Sigma_{\rm IR}}{10^{11}\,\lsun\,\kpc^{-2}}\right)^{-1/2}.
\end{equation}
Figure~\ref{fig:ratio} compares this result with {\it{Herschel}}/PACS observations of an FIR-selected sample of nearby LIRGs in which nuclear \citep{Diaz-Santos13} and extended \citep{Diaz-Santos14} emission are considered separately.  Additionally, \citet{Rosenberg15} use {\it{Herschel}} measurements to sort a sample of even brighter systems into three `classes' based on ratios of their CO line luminosities.  By comparing the average CO spectral line energy distributions of each group to the model in \citet{NK14}, we derive its characteristic surface star formation rate and IR surface brightness and plot their group-averaged [CII]/FIR ratios.

The agreement is excellent between the data and equation~\ref{eq:ciidef} with $\fcii=0.13$.  Our predicted slope is only slightly steeper than the fitting formula derived in \citet{Diaz-Santos14} and fully consistent with the measurements for both LIRGs and ULIRGs across nearly four orders-of-magnitude in IR surface brightness.  Moreover, comparing our model to the observations (and ignoring any other effects that may additionally be at work) places an additional constraint on $\fcii$ (beyond those described in \S\ref{sec:fcii}), of $0.1\lesssim \fcii \lesssim 0.17$.

Combining equations~\ref{eq:SIR}, \ref{eq:ciidef}, and~\ref{eq:Tdust} additionally connects the [CII] deficit to the average dust temperature: 
\begin{equation}\label{eq:ciidef1}
\frac{\lcii}{\lfir} \approx 2.2\times10^{-3}\,\frac{\fcii}{0.13}\,\left(\frac{\Td}{30\,\K}\right)^{-2}.
\end{equation}
Further translating equation~\ref{eq:ciidef1} into a relationship between [CII]/FIR and $S_{\rm 60\,\microns}/S_{\rm 100\,\microns}$, which parameterizes the FIR spectral energy distribution, is complicated \citep[e.g.,][]{DL07}.  However, recent work by \citet{Gullberg15} plots South Pole Telescope observations as a function of $\Td$, which they derive from fitting all 7 bands of their SED data with a modified black body \citep[see][]{Greve12}.  Figure~\ref{fig:ratio1} compares equation~\ref{eq:ciidef1} to the \citet{Gullberg15} results and to an unpublished comparison sample of low-redshift Great Observatories All-Sky Survey (GOALS) galaxies (Graci\'{a}-Carpio et al., in preparation), for which \citet{Gullberg15} also compute dust temperatures (though only using $S_{\rm 60\,\microns}$ and $S_{\rm 100\,\microns}$).  The agreement is once again excellent, suggesting that the connection of the dust temperature to the [CII] deficit is as a large-scale quantity correlated to the specific FIR emission (see \S\ref{sec:fir}).

\section{Other Explanations of the [CII] Deficit}\label{sec:alt}

Several other explanations for the [CII] deficit appear in the literature.  Here, we briefly discuss and compare to high-temperature saturation the two main alternatives: `dust-bounded' HII regions (\S\ref{sec:alt:dbhii}) and positively-charged dust grains (\S\ref{sec:alt:pcdg}).  Optically thick [CII] emission or additional FIR emission from active galactic nuclei have also been cited, but these are unlikely to be the main explanation for the deficit in most systems \citep[e.g.,][]{Luhman98, Malhotra01, Diaz-Santos14, Gullberg15}.  Of course, the primary advantage of the saturating [CII] mechanism is its robustness and simplicity.  It must be at work whether or not other effects are present in addition and, yet, manages to account for the entire observed deficit on its own.  Moreover, as exclusively local, PDR effects, the alternatives in this section miss the large-scale component of the [CII] deficit described in \S\ref{sec:deficit} and required to produce Figure~\ref{fig:ratio}.

\subsection{`Dust-bounded' HII Regions}\label{sec:alt:dbhii}

In the `dust-bounded' HII region theory \citep{Voit92, Bottorff98, Luhman03, Abel09, Gracia-Carpio11}, dust in highly-ionized gas absorbs a significant fraction of the UV and ionizing photons.  This results in both excess FIR emission from dust inside the HII region and reduced photoelectric heating from PAH molecules outside the HII region, which are starved of UV photons.  This scenario is a popular explanation for the [CII] deficit because of its impact on dust temperature, which results in a decreasing [CII]/FIR ratio with increasing ratio of $60\,\micron$ to $100\,\micron$ continuum flux \citep{Abel09}.  Moreover, it may help explain the behavior of the $9.7\,\micron$ silicate feature---which increases in absorption strength with decreasing [CII]/FIR ratio---by creating a larger temperature disparity between dust inside and outside HII regions \citep{Diaz-Santos13}.

On the other hand, even for large dust optical depths in an HII region, a significant fraction of Lyman-limit photons photo-ionize hydrogen instead of being intercepted by dust \citep{Draine11b}.  Moreover, radiation pressure can push dust out of giant HII regions around compact star clusters on short time scales if the gas is sufficiently dense; for a cluster of $10^{3}$ O stars and an rms density of $n_{\rm rms}=10^{3}\,\cm^{-3}$, the dust drift timescale is only $2\times10^{5}\,\yr$ \citep{Draine11b}.

Finally, the dust need not be significantly warmer than expected to explain the relationship between [CII]/FIR and $S_{\rm 60\,\microns}/S_{\rm 100\,\microns}$.  Figure~\ref{fig:ratio1} shows that the warmer effective dust temperatures associated with galaxies of higher average surface star formation rates are sufficient to reproduce the observations.  Thus, neither the enhanced FIR emission from HII regions nor the warmer dust temperatures provided by the `dust-bounded' HII region model are required by the data.

\subsection{Positively-Charged Dust Grains}\label{sec:alt:pcdg}

Another oft-invoked mechanism to explain the [CII] deficit is saturation of the photoelectric heating rate due to positive charging of dust grains, while [CII] remains an otherwise reliable calorimeter \citep{TH85, Hollenbach91, Malhotra97, Kaufman99, GC01, Wolfire03, Stacey10}.  To describe this quantitatively, consider that, in the limit where the grain charging is significant and $T<10^{4}\,\K$, the photoelectric heating rate is proportional to $G'^{.27}\,\ngas^{0.73}\,T_{\rm gas}^{-0.37}$ as opposed to just $G'$ in the case where grain charging has no effect.

As an illustrative exercise, consider a comparison between galaxies with IR surface brightnesses of $\Sigma_{\rm IR}=10^{10}\,\lsun\,\kpc^{-2}$ and those with $\Sigma_{\rm IR}=10^{12}\,\lsun\,\kpc^{-2}$, corresponding roughly to LIRGs and ULIRGs.  If dust in the [CII] emitting gas is heated by a local star cluster, then $G'$ is independent of average galaxy properties.  However, if the gas is exposed to an average interstellar background, then $G'$ is proportional to the surface star formation rate and the IR surface brightness.  Gas density should also scale as some function of $\ssfr\propto\Sg^2$; assume $10^{3}\,\cm^{-3}$ and $10^{4}\,\cm^{-3}$ for our low- and high-$\Sigma_{\rm IR}$ cases, respectively.  On the other hand, the gas temperature should be comparatively similar between the two groups, since whatever line dominates the PDR cooling is likely very temperature-sensitive, allowing it to compensate for any difference in the heating rate with only a small change in $T_{\rm gas}$.  Combining these rough estimates, the heating rate in the high-$\Sigma_{\rm IR}$ galaxies is about a factor of 15 (or about 5 if PDR gas is heated by a the local star cluster) higher than that in the low-$\Sigma_{\rm IR}$ galaxies despite an increasing the surface star formation rate by a factor of 100.

However, in the regime when the photoelectric efficiency is reduced by grain charging, the [CII] emission is already nearly saturated.  This is evident from the case considered in \S\ref{sec:cii}, in which conservative values of $\ngas=10^{4}\,\cm^{-3}$ and $G'=100$ produce an equilibrium gas temperature of $T_{\rm gas} \approx 106\,\K$.  Using these values, the second term in the denominator of equation~\ref{eq:Hpe} has a value of about 1.4, indicating that grain charging is only just starting to become important.  Moreover, despite the reduced photoelectric efficiency, the heating rate still increases with $\ssfr$.  Yet, Figure~\ref{fig:ratio} shows that the observed slope of the [CII]/FIR--$\Sigma_{\rm IR}$ relation is consistent with no additional impact on the specific [CII] luminosity at higher $\ssfr$.  Finally, while grain-charging is a robust mechanism for reducing the photoelectric efficiency, it is not clear that photoelectric heating is always the dominant contribution to the heating in PDR gas.  For example, thermalization of free electrons produced during ionizations of the hydrogen gas by cosmic rays and hard x-rays may be more important, particularly if the photoelectric efficiency is reduced \citep[see][section B1.1 and references therein]{Krumholz14a}.  Note that this process also results in a heating rate proportional to $\ssfr$.

On the other hand, grain charging could play an important role in producing a deficit of other FIR emission lines, especially of [OI]$63\,\micron$, which saturates at much higher temperatures and densities than [CII].  Such a mechanism might explain similarities among the suppression of these lines in a handful of systems with high FIR/CO(1--0) ratios \citep{Gracia-Carpio11}.

\section{Conclusions}\label{sec:conc}

The decline in the [CII]/FIR ratio in LIRGs and ULIRGs with increasing IR surface brightness is simply a consequence of the quantum mechanics of the upper fine-structure energy state in CII, which saturates at gas temperatures much larger than $91\,\K$, the temperature corresponding to the energy separation between the upper and lower states.  At higher temperatures, the upper state is not increasingly populated, and so the specific [CII] luminosity remains constant as both the specific FIR luminosity and the IR surface brightness increase, with $L_{\rm FIR}/\mgas \propto \Sigma_{\rm IR}^{1/2}$ resulting from their relative dependences on surface star formation rate and surface density.  This mechanism is a quantum mechanical necessity and must be at work independent of other possible contributions to the [CII] deficit, such as a diminished efficiency for photoelectic heating from positively charged dust grains or excess FIR emission and reduced heating as a result of `dust-bounded' HII regions.

While the slope of the [CII]/FIR--$\Sigma_{\rm IR}$ relation is set by [CII] saturation and the star formation law, the normalization reveals that about 10--17$\%$ of the ISM in LIRGs and ULIRGs is dissociated, [CII]-traced gas.  This fraction is roughly constant with surface density and consistent with observations of [CII]/CO(1--0), which probe the ratio of atomic to molecular gas in the ISM.  However, this value is an order-of-magnitude larger than the prediction from models that assume interstellar gas is configured as discrete molecular clouds bathed in an external radiation field that dissociates gas from the outside-in.  On the other hand, assuming that the structure of the molecular gas more closely resembles `swiss cheese,' with relatively disconnected PDR `holes' surrounding star clusters may represent a way forward for future calculations.  These contrasting configurations likely have different signatures in the [OI]$63\,\micron$ line as well as important implications for the escape of ionizing photons into the intergalactic medium.

\section*{Acknowledgements}
We thank Joaquin Vieira and Javier Graci\'{a}-Carpio for providing the low-redshift GOALS data and the derived corresponding dust temperatures.

\begin{appendix}
\section{Photoelectric Heating}\label{sec:app:heat}
The approximate rate of photoelectric heating from the dust, in which incident UV photons knock off electrons that subsequently thermalize with the gas, appears in equation~C5 from the appendix of \citet{Wolfire03} as
\begin{equation}\label{eq:Hpe}
\Gamma_{\rm pe}=\frac{1.1\times10^{-25}\,G'\,Z_{\rm d}}{1+3.2\times10^{-2}\,\left(G'\,T_{2}^{1/2}\,n_{e}^{-1}\,\cm^{-3}\,\phi_{\rm PAH}\right)^{0.73}} \,\erg\,\s^{-1}
\end{equation}
in the limit of $T_{\rm gas}<10^{4}$, where $G'$ is the FUV radiation field normalized to $G'_{\rm local}\approx 2.7\times10^{-3}\,\erg\,\s^{-1}\,\cm^{-2}$ (the value of the interstellar field in the solar neighborhood); $Z_{\rm d}$ is the dust-to-gas ratio normalized to the Galactic value; $T_{2}=T_{\rm gas}/(100\,\K)$; $n_{e}$ is the electron density, which we take to be $n_{e}\approx ({\rm C/H})\,n$; and $\phi_{\rm PAH}$ is a parameter related to PAH collision rates with a value of about 0.5.  In equation~\ref{eq:Hpe}, the second term in the denominator quantifies the reduced heating efficiency due to positively-charged dust grains.

\end{appendix}


\end{document}